\documentclass[9pt,twocolumn,twoside]{ownjnl}
\usepackage{txfonts}

\DeclareMathOperator{\erf}{erf}
\title{Unbiased centroiding of point targets close to the Cramer Rao limit}
\author[1]{Gerald Hechenblaikner}
\affil[1]{Airbus Space Systems, Airbus Defence and Space GmbH, Claude-Dornier-Straße, 88090 Immenstaad am Bodensee, Germany}
\begin{abstract}
Systematic errors affecting center-of-gravity (CoG) measurements may occur from coarse sampling of the point-spread-function (PSF) or from signal truncation at the boundaries of the region-of-interest (ROI).
For small ROI and PSF widths, these effects are shown to become dominant, but this can be mitigated by introducing novel unbiased estimators that are largely free of systematic error and perform particularly well for low photon number. Analytical expressions for the estimator variances, comprising contributions from photon shot noise, random pixel noise and residual systematic error, are derived and verified by Monte Carlo simulations. The accuracy and computational speed of the unbiased estimators is compared to those of other common estimators, including iteratively weighted CoG, thresholded CoG, iterative least squares fitting, and two-dimensional Gaussian regression. Each estimator is optimized with respect to ROI size and PSF radius and its error compared to the theoretical limit defined by the Cramer Rao lower bound (CRLB). The unbiased estimator with full systematic error correction operating on a small ROI [3x3] emerges as one of the most accurate estimators while requiring significantly less computing effort than alternative algorithms.
\end{abstract}

\begin{document}

\maketitle
\section{Introduction}
\begin{figure}
\centerline{\includegraphics[width=0.85\columnwidth]{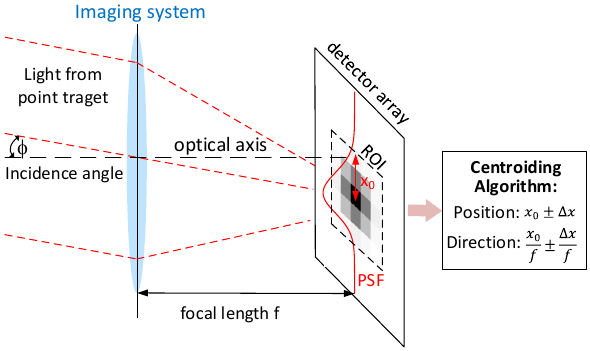}}
\caption  {Schematic of the measurement system: Light from a point target is incident at angle $\phi$ and focused by the imaging system to yield the PSF (red solid curve) which is converted into gray values by the pixels of the matrix sensor. The angle $\phi$ is given by the ratio of centroid displacement $x_0$ and focal length $f$.\label{fig_1}}
\end{figure}

Centroiding of point targets has been a focus of intense research since the early days of image sensors. 
It is a key aspect in many areas of instrumentation, ranging from attitude determination through star trackers \cite{liebe2002accuracy}, particle imaging velocimetry (PIV) \cite{wernet1993particle}, adaptive optics \cite{thomas2006comparison, baker2007iteratively} and astrometry \cite{lindegren2013high}, to breakthrough discoveries with fluorescence microscopy in the biomedical sciences \cite{hell2009microscopy, thompson2002precise}. Below, a summary of relevant work is given which is not meant to be extensive but limited to certain aspects in relation to centroiding performance.\\
A profound analysis of centroiding performance is already found in very early work by Grossman et al. \cite{grossman1984performance} who systematically investigated the dependency of Center-of-Gravity (CoG) centroiding on the pixel width and the size of the ROI.  Shortly afterwards Winick \cite{winick1986cramer} derived an expression for the minimum variance of centroid estimators (Cramer Rao Lower Bound) in presence of photon shot noise and pixel noise which was fundamental to understanding the ultimate limitations of centroiding. 
In the context of PIV analyses, Wernet and Pline \cite{wernet1993particle} showed that the measurement error for large particles scales linearly with particle size if accounting for photon noise only, which was confirmed to apply to the case of pixel noise as well by Westerweel  \cite{westerweel2000theoretical} who also showed the error to scale linearly with particle displacement.
Alexander et al. \cite{alexander1991elimination} investigated the systematic error incurred when the point-spread-function (PSF) is sampled with only few points through a Fourier space analysis. This also motivated further research by Jia et al. into how sampling errors can be mitigated through an estimator with bias correction\cite{jia2010minimum}, where photon shot noise only was considered as a source of random error. The Fourier space analysis was extended to include effects from finite size ROI in \cite{jia2010systematic}, where an approximate expression for the systematic error due to truncation effects was derived and a method for compensation suggested. However, the authors had not included the errors from random pixel noise or photon shot noise in this assessment.
An analytical derivation of the CoG error variance due to random pixel noise and an approximation of the variance due to photon shot noise was presented bv Cao et al. in \cite{cao1994accuracy}, which was extended to the significantly more complex case of thresholded CoG by Ares et al. in \cite{ares2004influence}. However, the latter analysis did not include photon noise and is limited to a given static intensity distribution, while in general applications the target image is expected to be located randomly on the focal plane, thus requiring an ensemble average over possible locations to determine a representative error distribution. 
The dependency of the centroid variance of a diffraction-limited PSF on the size of the ROI is investigated in \cite{irwan1999analysis}, where only photon shot noise was considered. 
Weighted CoG (WCoG) methods \cite{nicolle2004improvement} apply a weighting function to the intensity distribution, thereby suppressing the error incurred from noisy pixels at the distribution tails and beyond. Analytical predictions of the error variance were derived by Nicolle et al. \cite{nicolle2004improvement} and corrected by Thomas et al. \cite{thomas2006comparison} who investigated the WCoG performance in detail in comparison to thresholded CoG and correlation methods. However, WCoG can only be applied if the distribution center is already sufficiently well known in order to track small excursions from it. While this helps for accurate tracking, it is unsuitable for initial target acquisition. This problem is overcome by iteratively weighted CoG (IWCoG) methods, where the center of gravity obtained for a certain iteration step is used to adjust the center of the weighting function for the next iteration step \cite{baker2007iteratively, akondi2010improved}.
Bao et al. adopted an energy iteration algorithm for spot target localization and found significant improvements in terms of accuracy and speed with respect to thresholded CoG \cite{bao2022window}. Similarly, consecutive adjustments of ROI were shown by Gao. et al. to improve centroiding accuracy at very low light levels \cite{gao2020high}.
Using a Kalman filter, Sun et al. \cite{sun2022centroid} could largely suppress the periodic systematic error which occurs when tracking a dim point target.\\
This paper investigates the achievable limits for centroiding of point targets, i.e., objects whose image size is dominated by the PSF. 
The main aims and novel aspects are: (1) rigorous theoretical derivation of all contributions to the CoG error variance, including systematic errors (sampling and truncation errors), photon shot noise, and pixel noise; (2) definition and assessment of unbiased estimators capable of removing systematic errors, making it favorable to use small ROIs and PSF radii; (3) performance comparison between the novel unbiased estimators and other estimators with respect to centroiding accuracy and computational effort for different ROI sizes and photon numbers.\\
The paper is structured as follows:
In section \ref{sec:Overview} a thorough investigation of systematic errors is performed and novel unbiased estimators are introduced. Section \ref{sec:ROI_impact} derives the probability density function (PDF) for systematic error and investigates the effect of error correction in presence of photon and pixel noise for different (small) ROI sizes and photon numbers. Estimator variances for linear and full systematic error correction are derived in section \ref{sec:variance},  showing that unbiased estimators remove systematic error but broaden the error contributions from photon and pixel noise.
In the concluding section \ref{sec:other_centroiding_methods}, the unbiased estimators are compared to some common estimators, including thresholded CoG, iteratively weighted CoG, iterative least squares fitting, and two-dimensional Gaussian regression. For each estimator, the optimal ROI size and PSF radius are found and the performances compared to the CRLB. Optimization is performed with respect to the normalized centroid error, expressed as a fraction of the PSF radius, which directly relates to the achievable angular accuracy. Results indicate that the unbiased estimator with full systematic error correction and a small ROI [3x3] is among the best performing estimators and comes very close to the CRLB while requiring significantly less computing effort.\\
The system model assessed in this paper consists of a matrix detector with associated read-out-noise and dark current, and an imaging system of a variable focal length $f$ that determines the PSF size on the detector, as shown in Fig.~\ref{fig_1}. 

\section{Overview of systematic errors}
\label{sec:Overview}
The main contributors to systematic error are due to (1) finite sampling of the PSF width (sampling error) and (2) truncation of the intensity distribution at the boundaries of the ROI (truncation error). The derivation of the sampling error in section \ref{sec:sampling_error} is based upon the Fourier space analysis approach of Alexander \cite{alexander1991elimination} which is extended to account for finite ROI size in section \ref{sec:truncation_error}.
\subsection{Sampling error}
\label{sec:sampling_error}
It is assumed that light with a Gaussian-shaped intensity profile $I_s(x,y)$ of radius $\sigma_g$ and centered around $(x_0,y_0)$ is incident on the detector array:
\begin{eqnarray}
I_s(x,y,x_0,y_0)&=&\frac{1}{2\pi\sigma_g^2}e^{-\frac{(x-x_0)^2+(y-y_0)^2}{2\sigma_g^2}}\label{eq:Gaussian2d}\\
I_s(x,x_0)&=&\frac{1}{\sqrt{2\pi}\sigma_g}e^{-\frac{(x-x_0)^2}{2\sigma_g^2}}\label{eq:Gaussian1d}
\end{eqnarray}
Considering that the 2-dimensional (2d) Gaussian separates in the product of two 1d Gaussian functions, the following analysis is based on the 1d case of Eq.~\ref{eq:Gaussian1d} and can easily be generalized to the 2d-case. The incident light intensity is integrated by the pixel response function $p(x)$, which is assumed to be a box-car function of pixel length $\Delta$:
\begin{equation}
p(x)=
\begin{array}{l}
1,~|x|\le \Delta/2\\
0,~|x|> \Delta/2.
\end{array}
\label{eq:pixel_function}
\end{equation}
The resulting pixel-broadened distribution $f(x)$ is obtained by the convolution of $I_s(x)$ with $p(x)$:
\begin{eqnarray}
f(x,x_0)=(I_s*p)(x)=\int_{-\infty}^{\infty} I_s(x',x_0)p(x'-x)dx'&&\nonumber\\
=\frac{1}{2\Delta}\left[\erf\left(\frac{x-x_0+\frac{\Delta}{2}}{\sqrt{2}\Delta}\right)-\erf\left(\frac{x-x_0-\frac{\Delta}{2}}{\sqrt{2}\Delta}\right)\right] &&\label{eq:f_broadened}
\end{eqnarray}
Then $f(x,x_0)$ is multiplied by the sampling function $s(x)$, which is represented by a comb of delta-peaks with a spacing of the pixel length $\Delta$,
\begin{equation}
s(x)=\sum_{n=-\infty}^{\infty}\delta(x-n\Delta)
\label{eq:sampling_function}
\end{equation}
to obtain the sampled intensity $g(x)$
\begin{equation}
g(x,x_0)=f(x,x_0)\cdot s(x)=(I_s*p)(x)\cdot s(x)
\label{eq:g_x}
\end{equation}
The Center-of-Gravity (CoG) $X_c$ is defined as the first moment of the sampled intensity function:
\begin{equation}
X_{c}=\frac{\int_{-\infty}^{\infty}xg(x,x_0) dx}{\int_{-\infty}^{\infty}g(x,x_0) dx}=\frac{\sum_{i} x_i g_i}{\sum_{i}g_i},
\label{eq:g_cog}
\end{equation}
where the $x_i$ denote the coordinates of the pixel centers $x_i\in\{..,-1,0,1,..\}$. $X_c$ can also be found from the Fourier transform $\tilde{G}(s)$ of $g(x,x_0)$ through the relation \cite{alexander1991elimination}:
\begin{equation}
X_{c}=\frac{\tilde{G}'(0)}{-i2\pi \tilde{G}(0)}.
\label{eq:CoG_Fourier}
\end{equation}
Note that $\tilde{G}(s)$ can be expressed in terms of the Fourier transforms of $I_s(x,x_0)$, $p(x)$, and $s(x)$, which are denoted as $\tilde{I}_s(s)$, $\tilde{P}(s)$, and $\tilde{S}(s)$, respectively, as follows: 
\begin{equation}
\tilde{G}(s)=\tilde{F}(s)*\tilde{S}(s)=(\tilde{I}(s)\tilde{P}(s))*\tilde{S}(s)\label{eq:G_s_short}
\end{equation}
When $f(x,x_0)$ is Fourier-transformed to obtain $\tilde{F}(s)$, it is useful to remove the explicit dependency on $x_0$ by introducing the even function $f_e(x)=I_s(x,0)*p(x)$ and its Fourier transform $\tilde{F}_e(s)$. Using the Fourier shift theorem and calculating the Fourier transforms of Gaussian and boxcar functions, $\tilde{F}(s)$ is found to be:
\begin{equation}
\tilde{F}(s)=e^{-i2\pi s x_0}\tilde{F}_e(s)=e^{-i2\pi s x_0}e^{-2\pi^{2}s^{2}\sigma_g^{2}}\frac{\sin\left(\pi s\Delta\right)}{\pi s\Delta}.
\label{eq:F_s}
\end{equation}
Noting that the Fourier transform of the comb function $s(x)$ is itself a comb function
\begin{equation}
\tilde{S}(s)=\sum_{n=-\infty}^{\infty}\frac{1}{\Delta}\delta(s-\frac{n}{\Delta}),
\label{eq:FT_s}
\end{equation}
the following expressions for $\tilde{G}(s)$ and its derivative are obtained from Eqs.~\ref{eq:G_s_short},\ref{eq:F_s},\ref{eq:FT_s}:
\begin{eqnarray}
\tilde{G}(s)&=&\frac{1}{\Delta}\sum_{n=-\infty}^{\infty}e^{-i2\pi x_{0}(s-\frac{n}{\Delta})}\tilde{F}_{e}(s-\frac{n}{\Delta})\label{eq:G_s_full}\\
\tilde{G}'(s)&=&\frac{1}{\Delta}\sum_{n=-\infty}^{\infty}\left[\left(-i2\pi x_{0}\right)e^{-i2\pi x_{0}(s-\frac{n}{\Delta})}\tilde{F}_{e}(s-\frac{n}{\Delta})\right.\nonumber\\
						 &+& \left.e^{-i2\pi x_{0}(s-\frac{n}{\Delta})}\tilde{F}'_{e}(s-\frac{n}{\Delta})\right].\label{eq:G'_s_full}
\end{eqnarray}
Evaluating $\tilde{F}_e(s)$ of Eq.~\ref{eq:F_s} and its derivative for $s=0$, and substituting results into Eqs.~\ref{eq:G_s_full},\ref{eq:G'_s_full} yields:
\begin{align}
\tilde{G}(0)&=\frac{1}{\Delta}\tilde{F}_{e}(0)=\frac{1}{\Delta}\\
\tilde{G}'(0)&=\frac{-i2\pi}{\Delta}\left[ x_{0}+\frac{1}{\pi}\sum_{n=1}^{\infty}\sin\left(2\pi x_{0}\frac{n}{\Delta}\right)e^{-2\pi^{2}\left(\frac{n}{\Delta}\right)^{2}\sigma^{2}}\frac{\Delta}{n}\left(-1\right)^{n}\right]
\end{align}
The sampling error $\delta_{x,samp}$ is the found from the difference between the centroid and the actual center $x_0$ of the intensity distribution:
\begin{align}
\delta_{x,samp}&=X_c-x_0=\frac{\tilde{G}'(0)}{-i2\pi \tilde{G}(0)}-x0\\
&=\frac{1}{\pi}\sum_{n=1}^{\infty}\sin\left(2\pi x_{0}\frac{n}{\Delta}\right)e^{-2\pi^{2}\left(\frac{n}{\Delta}\right)^{2}\sigma^{2}}\frac{\Delta}{n}\left(-1\right)^{n}
\label{eq:sampling_error}
\end{align}
Note that for any practical application only the first term of the sum in Eq.~\ref{eq:sampling_error} is relevant, as higher order terms are much smaller and can generally be neglected.

\subsection{Truncation error}
\label{sec:truncation_error}
However, truncation of the intensity due to the finite size of the ROI must also be considered, which is achieved by multiplying the sample intensity $g(x,x_0)$ of Eq.~\ref{eq:g_x} by a window function $w(x)$, where $N_s$ corresponds to the ROI width in pixels, to obtain the windowed sampled intensity $g_w(x)$:
\begin{align}
w(x,N_s) &=
\begin{array}{cc}
1,\hspace{1em} |x|\le N_s\Delta/2\\
0,\hspace{1em} |x|> N_s\Delta/2
\end{array}\\
g_w(x,x_0)&= (I*p)(x)\cdot w(x)\cdot s(x)
\label{eq:g_w}
\end{align}
In order to determine the centroid according to Eq.~\ref{eq:CoG_Fourier}, the Fourier transform of $g_w(x,x_0)$ must be obtained from Eq.~\ref{eq:g_w}:
\begin{equation}
\tilde{G}_w(s)=(\tilde{I}(s)\cdot \tilde{P}(s))*\tilde{W}(s)*\tilde{S}(s)\label{eq:wind_meth_fourier}.
\end{equation}
Equation~\ref{eq:wind_meth_fourier} describes the approach taken in \cite{jia2010systematic}, but the computations are quite elaborate and approximations in terms of Taylor expansion must be made in order to derive an analytical expression for the limit where truncation errors dominate over sampling errors.\\
A simple approximation for the truncation error is obtained by performing the centroid using the continuous shape profile of the incident light $I_s(x,x_0)$ instead of using the discrete sample intensity function $g(x,x_0)$, which is valid in the limit $\sigma/\Delta \gg 1$. For ease of notation, all parameters are expressed in units of the pixel length $\Delta$ in the following derivations. The centroid is then found to be
\begin{align}
X_c\approx&\frac{\int_{-N_s/2}^{-N_s/2} xe^{\frac{(x-x_0)^2}{2\sigma_g^2}}dx}{\int_{-N_s/2}^{-N_s/2} e^{\frac{(x-x_0)^2}{2\sigma_g^2}}dx}\\
=&x_0-\sqrt{\frac{2}{\pi}}\sigma_g
\frac{e^{\frac{-\left(N_s/2-x_0\right)^2}{2\sigma_g^2}}-e^{\frac{-\left(N_s/2+x_0\right)^2}{2\sigma_g^2}}}{\erf\left(\frac{N_s/2-x_0}{\sqrt{2}\sigma_g}\right)+\erf\left(\frac{N_s/2+x_0}{\sqrt{2}\sigma_g}\right)}.
\label{eq:truncation_error}
\end{align}
In the limit of small displacements $x_0$ from the center pixel Eq.~\ref{eq:truncation_error} can be approximated by Taylor expansion up to first order in $x_0$: 
\begin{eqnarray}
\delta_{x,cut}&=&X_c-x_0=-x_{0}\sqrt{\frac{2}{\pi}}\frac{N_s}{2\sigma_g}\frac{e^{-\frac{N_s^2}{8\sigma_g^2}}}{\erf\left(\frac{N_s}{2\sqrt{2}\sigma_g}\right)} +O(x_0^3)\nonumber\\
&=&x_0F_{cut},\label{eq:linear_truncation_error}
\end{eqnarray}
where the negative constant factor $F_{cut}$ was introduced that relates the truncation error to the displacement $x_0$.
In fact, Eq.~\ref{eq:linear_truncation_error} is valid to zeroth order in $1/\sigma_g^2$. Taylor expansion up to first order in $1/\sigma_g^2$ of the total systematic error yields an additional corrective factor of $[1+1/(12\sigma_g^2)]$, as obtained from the ansatz given by Eq.~\ref{eq:wind_meth_fourier} in \cite{jia2010systematic}.

\subsection{Total systematic error}
\begin{figure}
\centerline{\includegraphics[width=0.85\columnwidth]{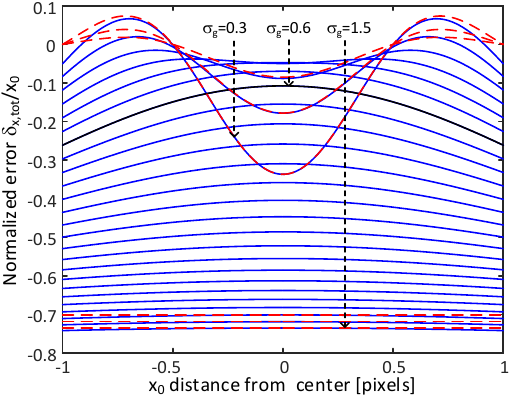}}
\caption  {Normalized total systematic error $\delta_{x,tot}/x_0$ plotted against displacement $x_0$ of the intensity profile $I_s(x)$ from ROI center for ROI [3x3]. The individual blue lines correspond to different PSF radii $\sigma_g$, varying from 0.3 to 1.5 pixels in steps of 0.05 pixels. Black line indicates plot for $\sigma_g=0.6$. Analytical approximations for small and large $\sigma_g$ are given by the top and bottom red-dashed lines, respectively.\label{fig_2}}
\end{figure}

In general, the total systematic error comprises both, sampling and truncation error and neither of the two limits described above applies.
However, it is easy to compute the total systematic error as a function of the displacement $x_0$ as follows:
\begin{equation}
\delta_{x,tot}(x_0)=\frac{\sum_i x_i f(x_i,x_0)}{\sum_i f(x_i,x_0)}-x_0\label{eq:total_systematic},
\end{equation}
where the pixel-broadened function $f(x,x_0)$ of Eq.~\ref{eq:f_broadened} is evaluated at the pixel centers $x_i$ for a given value of $x_0$.\\
Note that the total systematic error $\delta_{x,tot}$ is a monotonous function in $x_0$ and rises approximately linearly with $x_0$ for large PSF radii.
For plotting, it is therefore convenient to normalize it by $x_0$ , as shown in Fig.~\ref{fig_2}, where $\delta_{x,tot}/x_0$ is plotted against the displacement $x_0$ of $I_s(x)$ from the ROI [3x3] center (blue lines). The individual curves correspond to different values of $\sigma_g$. The normalized error oscillates for small $\sigma_g$, where the sampling error dominates and the analytical approximation of Eq.~\ref{eq:sampling_error} applies. The top red-dashed lines, indicating the analytical predictions for $\sigma_g=0.3, 0.35, 0.40$, agree very well with the exact numerical results. For $\sigma_g\ge 0.5$ the oscillation disappears and the normalized error peaks in the center, gradually decreasing with larger values of $\sigma_g$.\\
The normalized error approaches a flat line for large values of $\sigma_g$, where the truncation error dominates and the analytical approximation of Eq.~\ref{eq:linear_truncation_error} (including corrective term) applies. The bottom red lines designate analytical predictions for $\sigma_g=1.4, 1.45, 1.50$ and match the numerical results very well.
Knowledge of the total systematic error allows calculating the error variance. Assuming that the intensity distribution offset $X_0$ is uniformly distributed around the ROI center pixel in the interval $[-0.5,0.5]$, the variance can be easily calculated by:
\begin{equation}
\sigma_{sys}^2=\mathrm{Var}(X_c-X_0)=\int_{-0.5}^{0.5} \delta^2_{x,tot}(x_0) dx_0.\label{eq:variance_total_systematic_error}
\end{equation}
A similar expression was first suggested by Jia \cite {jia2010minimum} for the periodic sampling error, given by Eq.~\ref{eq:sampling_error}. However, it can just as well be applied to the total systematic error of Eq.~\ref{eq:total_systematic} or to the linearized error of Eq.~\ref{eq:truncation_error}. The latter dominates for small ROI and relatively large $\sigma_g$, where Eq.~\ref{eq:linear_truncation_error} implies that the truncation error relates to the displacement $x_0$ via a constant (negative) factor $F_{cut}$, as represented by the bottom curves of Fig.~\ref{fig_2}. Consequently, the linearized error is uniformly distributed between $[F_{cut}/2,-F_{cut}/2]$.\\
More generally, the probability density function (PDF) of the total systematic error, represented by a random variable $Y$,  can be calculated. Consider that $Y$ relates to the uniformly distributed random variable $X_0$, representing the displacement of the intensity distribution, through $Y=\delta_{x,tot}(X_0)$. Then, noting that the cumulative distribution function (CDF) of $X_0$ is given by $CDF_{X_0}(x_0)=0.5+x_0$, the CDF and PDF of $Y$ are obtained as follows:
\begin{eqnarray}
CDF_Y(y)&=&\mathrm{P}\{Y<y\}=\mathrm{P}\{X_0<\delta^{-1}_{x,tot}(y)\}\nonumber\\
&=&CDF_{X_0}(\delta^{-1}_{x,tot}(y))=0.5+\delta^{-1}_{x,tot}(y)\nonumber\\
PDF_Y(y)&=&\frac{d}{dy}CDF_Y(y)=\frac{d}{dy}\delta^{-1}_{x,tot}(y),
\label{eq:PDF_sys_err}
\end{eqnarray}
where $PDF_Y(y)$ is defined between the interval bounds of $[\delta_{x,tot}(0.5),\delta_{x,tot}(-0.5)]$ and the function $\delta_{x,tot}(x_0)$ is assumed to strictly monotonic decrease in between $[-0.5,0.5]$. 
The latter condition is met for values of $\sigma_g\ge 0.43$. Below this threshold, sampling errors dominate and affect an oscillatory, i.e. non-monotonic,  behavior of $\delta_{x,tot}(x_0)$. In this case, the probability density function bi-furcates and broadens abruptly, leading to a sharp increase of the variance towards lower values of $\sigma_g$.\\
For all useful practical purposes, the PSF radius must be sufficiently large in order to avoid the rapid increase of variance, in which case $\delta_{x,tot}(x_0)$ is always strictly monotonic. Therefore, a simple pre-computed (or measured) lookup table is sufficient to establish the functional relationship $f_{lu}(x_0)$ between the actual position $x_0$ and the position $X_c$ determined for the centroid, which is obtained from Eq.~\ref{eq:total_systematic} as follows:
\begin{eqnarray}
X_c&=&x_0+\delta_{x,tot}(x_0)=f_{lu}(x_0)\nonumber\\
x_0&=& f^{-1}_{lu}(X_c)\label{eq:sys_err_correction}
\end{eqnarray}
Therefore, the following unbiased estimator $X_{c,ub}$ corrects the systematic error in the CoG measurement represented by $X_c$
\begin{equation}
X_{c,ub}=f^{-1}_{lu}(X_c),
\label{eq:corrected_estimator}
\end{equation}
The work of \cite{jia2010minimum} investigated the impact of correcting the sampling error of Eq.~\ref{eq:sampling_error} alone, referred to as ``S-curve bias'', using an iterative numerical approach.
In this paper, the impact of correcting the full systematic error is assessed, which is necessary for small ROI, based on simple interpolation of a lookup table alone and without recourse to complex iterative calculations.
It is obvious that random noise contributions, such as pixel noise and photon noise, also affect the measured centroid position, which distorts the direct relationship given in Eq.~\ref{eq:sys_err_correction}. Its impact on centroid error is analyzed in section  \ref{sec:variance}, where an analytical expression for the variance of the estimator $X_{c,ub}$ is derived. 

\section{Impact of of ROI size on centroiding}
\label{sec:ROI_impact}
This section analyzes the impact of ROI size on centroiding error, considering various sources of random noise and systematic error.

\subsection{Noise sources}
\label{sec:noise_sources}
Monte Carlo (MC) simulations of the centroiding are performed, where additive Gaussian noise is used to simulate the impact of random pixel noise and a Poissonian distribution of the light intensity on each pixel accounts for average signal and photon (shot) noise. Pixel noise comprises contributions from the detector dark current, pixel read-out noise (RON), and quantization noise. Typical values of its standard deviation $\sigma_\eta$ may range from a few photoelectrons ($e^-$) to several hundred $e^-$. For our simulations, we assumed $\sigma_\eta=10\,e^-$ and typically performed a total number of $N_{MC}=2\times 10^4$ simulations, unless stated otherwise. For ease of discussion and without loss of generality, the detector quantum efficiency (fraction of photo-electrons per incident photon) is assumed to be 100\%, i.e. 1 photon corresponds to 1 $e^{-}$. Quantization noise, resulting from conversion of electrons to digital units, is generally small compared to other noise sources and omitted in the following discussions.\\
A useful quantity directly correlating with the achievable centroiding accuracy is the signal-to-noise ratio (SNR) within a given ROI of width $N_s$,  given by:
\begin{equation}
SNR=\frac{N_p'}{(N_s^2\sigma_\eta^2+N_p')^{1/2}},
\label{eq:SNR}
\end{equation} 
where $N_p'$ is the number of photoelectrons within the ROI which -accounting for truncation- relates to the total number of photoelectrons $N_p$ through the relation $N_p'=N_p\erf^2(N_s/(2\sqrt{2}\sigma_g))$.
Following the definition by Liebe \cite{liebe2002accuracy}, the detection threshold (lowest detectable signal level in units of $N_p$) is given by $I_{det}=20\sigma_\mu/\erf^2(1/\sqrt{2}\sigma_g)$.
In our simulations, $N_p$ typically ranged from $5\times 10^2$ to $5\times 10^4$, where the lower limit is close to the detection threshold $I_{det}\approx 430\,e^{-}$ for $\sigma_g=1$.

\subsection{Discussion of simulation results}
\begin{figure}
\centerline{\includegraphics[width=0.90\columnwidth]{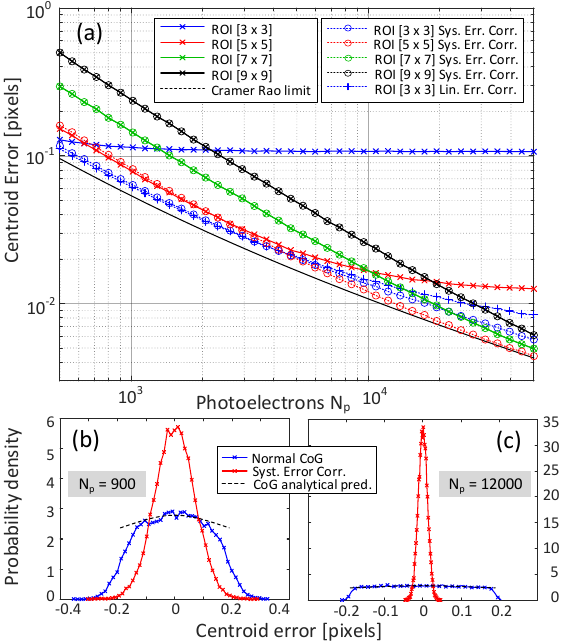}}
\caption  {(a) Simulated centroid error plotted against $N_p$. Colors designate ROI sizes of $N_s=3$ (blue), 5 (red), 7 (green) and 9 (black), respectively. x-marks 'x' refer to normal CoG, circles 'o' to CoG with systematic error correction, plus signs '+' to CoG with linear error correction. (b) PDF of simulated centroid errors for $N_p=900$, $N_s=3$, and $\sigma_g=0.85$; normal CoG (blue), CoG with systematic error correction (red), analytical PDF prediction for normal CoG (black dashed line) . (c) Same parameters as in (b) but $N_p=1.2\times 10^4$.\label{fig_3}}
\end{figure}

Figure \ref{fig_3}(a) plots the simulated centroid errors as a function of the number of photoelectrons $N_p$, where $N_p$ ranges from $500$ to $5\times 10^4$ and $\sigma_g=0.85$ (FWHM equals 2). The colors refer to results derived for different ROI widths, where widths of $N_s=3,5,7,9$ are represented by colors blue, red, green, and black, respectively. The solid lines designate errors obtained for conventional CoG centroids, the dashed lines refer to errors obtained from estimator $X_{c,ub}$ of Eq.~\ref{eq:corrected_estimator} which corrects the total systematic error. The fundamental limit of centroid accuracy, the Cramer Rao lower variance bound, is computed considering photon shot noise as well as pixel noise for a two-dimensional estimator, as outlined in the derivation by Winick \cite{winick1986cramer}, and plotted as the black dashed line. Additional aspects of CRLB position estimators in relation to Gaussian and Poissonian noise are discussed by Fischer in \cite{fischer2019limiting}.\\
It is apparent that the CoG error for ROI [3x3] (blue line with x-marks) does not decrease with $N_p$ but remains very high at 0.1 pixels. However, performing a systematic error correction (blue line with circles) yields a vastly improved performance, with the error decreasing down to less than 0.01 pixels for the highest value of $N_p$. Almost equally small centroiding errors are obtained when using the linear error correction of Eq.~\ref{eq:linear_truncation_error} (dashed blue line with marker '+') together with the corrective factor $[1+1/(12\sigma_g^2)]$, indicating that the systematic error is dominated by truncation effects. Both, total and linear, systematic error corrections yield consistently smaller errors than those of the uncorrected CoG for ROI [5x5] (red line with x-marks) which generally also shows good performance and is considerably less affected by systematic error due to its larger ROI size. For $N_p>10^4$, the ROI [5x5] centroid error does not decrease anymore because random errors have become so small that the systematic error dominates. This is overcome by systematic error correction, as indicated by the red-dashed line with circles, which approaches the Cramer Rao limit for high $N_p$. \\
While the centroid errors with error correction for ROI [3x3]  are smaller than those for ROI [5x5] at low $N_p$, this reverses for $N_p>5000$. The reason is that for this relatively large spot width, $15\%$ of the signal power are truncated for ROI [3x3] but much less than $0.1\%$ for other ROI sizes. In contrast, ROI [3x3] has the advantage of accumulating less pixel noise due to the smaller number of pixels. While this improves performance for low $N_p$, the benefit is lost for high $N_p$, where pixel noise becomes insignificant compared to shot noise, and the signal loss at the ROI edges leads to a worse performance of ROI [3x3].  However, this performance limitation can be mitigated by reducing the spot width to $\sigma_g=0.6$, as will be seen in section \ref{sec:discussing_variances}.

The centroid errors for larger ROIs [7x7] (green) and [9x9] (black) are not affected by systematic error at all so that corrected and uncorrected CoG yield exactly the same results. However, owing to their larger size, more pixel noise is accumulated across the ROI, leading to significantly worse performance than that of smaller ROIs for low and medium values of $N_p$. For comparison, the SNR values for ROI [3x3], as computed from Eq.~\ref{eq:SNR}, range from 12 ($N_p=500$) to 204 ($N_p=5\times 10^4$), while those for ROI [9x9] range from 5 to 207, indicating the advantage of ROI [3x3] for small numbers of photoelectrons. 

Figures \ref{fig_3}(b) and (c) plot the PDF of centroid errors for ROI [3x3] and for 900 and $1.2\times 10^4$ photoelectrons, respectively. It is apparent that the distribution deviates strongly from Gaussian and is well described by the PDF of systematic errors calculated from Eq.~\ref{eq:PDF_sys_err} (black dashed line) for both, small and large $N_p$. While this indicates the prevalence of systematic error over random noise, the latter is non-negligible for small $N_p$ (Fig.~\ref{fig_3}b) and affects a smear-out at the corners. In contrast, for large $N_p$ (Fig.~\ref{fig_3}c) the systematic errors are so dominant that the PDF for normal CoG errors (blue line) is many time broader than the one obtained after correction of the full systematic error (red line). An approximation for the PDF is found by noting that the normalized systematic error for $\sigma_g=0.85$ is almost constant across one pixel (see Fig.~\ref{fig_2}) and therefore defined by the truncation factor $F_{cut}$ of Eq.~\ref{eq:linear_truncation_error}. The ensuing PDF is uniformly distributed in between $[F_{cut}/2,-F_{cut}/2]=[-0.18,0.18]$.

\section{Calculating estimator variances}
\label{sec:variance}

\subsection{CoG Variance without systematic error}
The intensity of pixel (i,j) in the presence of noise, $I'_{ij}$, is assumed to be a random variable that is given by the sum of the mean pixel intensity $I_{ij}=\langle I'_{ij}\rangle$ , random pixel noise $\eta_{ij}$ and photon shot noise $p_{ij}$:
\begin{equation}
I'_{ij}=I_{ij}+\eta_{ij}+p_{ij}.
\end{equation}
The mean pixel intensities are found by extending the 1 dimensional shape function of Eq.~\ref{eq:f_broadened} to 2 dimensions and multiplying by the photon number
\begin{equation}
I_{ij}=N_p f(x_i,x_0)f(y_j,y_0).
\label{eq:intensity_sampling}
\end{equation} 
Here, the pixel center coordinates are given by $(x_i,y_j)\in\{-(N_s-1)/2,..,-1,0,1,...,(N_s-1)/2\}$. In an extension of Eq.~\ref{eq:g_cog} from 1 to 2 dimensions, the center of gravity is given by:
\begin{equation}
X_c=\frac{\sum_{ij}x_i I'_{ij}}{\sum_{ij}I'_{ij}}.
\label{eq:xc_with_noise}
\end{equation}
The sum over noisy pixel intensities in the denominator of Eq.~\ref{eq:xc_with_noise} can be approximated by the sum over mean intensities $\sum_{i,j}I'_{ij}=\sum_{ij}I_{ij}=I_{tot}$. This is a frequently used approximation \cite{ares2004influence,thomas2006comparison} that is valid if the standard deviation of the denominator, representing the total photon number within the ROI, is much smaller than its mean value, as applicable in most practical situations.
The variance of $X_c$ is then found from Eq.~\ref{eq:xc_with_noise} to be:
\begin{eqnarray}
\mathrm{Var}(X_c)&=&\frac{\langle(\sum_{i,j} x_i(I_{ij}+\eta_{ij}+p_{i,j})-\sum_{i,j}x_i I_{ij})^2\rangle}{I_{tot}^2}\nonumber\\
&=& \frac{\sum_{i,j}x_i^2\langle \eta_{ij}^2\rangle}{I_{tot}^2}+\frac{\sum_{i,j}x_i^2 I_{ij}}{I_{tot}^2},
\label{eq:cog_variances}
\end{eqnarray}
where brackets $\langle...\rangle$ refer to a statistical ensemble average. It was assumed that the random variables for pixel and photon noise are independent, $\langle \eta_{ij} p_{nm}\rangle=0$, and that the random variables between pixels are also independent, $\langle \eta_{ij}\eta_{nm}\rangle =\langle\eta_{ij}^2\rangle\delta_{in}\delta_{jm}$ and $\langle p_{ij}p_{nm}\rangle =\langle p_{ij}^2\rangle\delta_{in}\delta_{jm}$.
The first term in the last line of Eq.~\ref{eq:cog_variances} describes the impact of pixel noise with constant variance $\sigma_\eta^2=\langle\eta_{ij}^2\rangle$ and the second term the impact of photon noise with a variance given by the pixel intensity $I_{ij}$.\\
The centroid variance due to random pixel noise $\sigma_{pix}^2$ can be further simplified to obtain
\begin{eqnarray}
\sigma_{pix}^2 &=& \frac{\langle \hat{\eta}_{ij}^2\rangle}{I_{tot}^2}\sum_{i,j}x_i^2=\frac{\sigma_\eta^2}{I_{tot}^2}\frac{N_s^2(N_s^2-1)}{12}\nonumber\\
&=&\frac{\sigma_\eta^2}{N_p^2\erf^2(N_s/(2\sqrt{2}\sigma_g))}\frac{N_s^2(N_s^2-1)}{12}.
\label{eq:variance_pixel_noise}
\end{eqnarray}
Note that a similar result was first derived in \cite{cao1994accuracy}. It is modified here to account for truncation of the total photon number $N_p$ at the boundary of the ROI, which is described by the square of the error function in the denominator.\\
Note that $I_{ij}$ in Eq.~\ref{eq:cog_variances} explicitly depends on the center positions $(x_0,y_0)$ which are assumed to be uniformly distributed in the interval $[-0.5,0.5]$.
Therefore, strictly speaking, when the variance due to photon noise $\sigma^2_{phot}$ is computed, the pixel intensity distribution $I_{ij}$ should  be averaged for all possible values of $x_0,y_0$. 
However, to a very good approximation with errors $<5\%$ for $\sigma_g\geq0.6$, the pixel intensities may be evaluated for $x_0,y_0=0$ only instead of averaging. It then follows from Eq.~\ref{eq:cog_variances}:
\begin{equation}
\sigma_{phot}^2\approx\frac{\sum_{i,j}x_i^2 I_{ij}(x_0,y_0=0)}{I_{tot}^2}=\frac{\sum_{i,j}x_i^2 f(x_i,0)f(y_j,0)}{N_p\erf^2(N_s/(2\sqrt{2}\sigma_g))}
\label{eq:variance_shot_noise_sum}
\end{equation}
It is easy to numerically evaluate the summands in the numerator of Eq.~\ref{eq:variance_shot_noise_sum} and perform the sum over the full ROI. For fine sampling ($\sigma_g\gg 1$), the values $I_{ij}$ can be further approximated by direct sampling of the intensity shape function $I_{ij} \approx I_s(x_i,y_i,0,0)$ of Eq.~\ref{eq:Gaussian2d}, and replacing the discrete sum by an integral to obtain:
\begin{align}
\sigma_{phot}^2&\approx \frac{N_s}{I_{tot}^2}\int_{-N_s/2}^{N_s/2}x^2 e^{-\frac{x^2}{2\sigma_g^2}}dx\int_{N_s/2}^{N_s/2}e^{-\frac{y^2}{2\sigma_y^2}}dy\nonumber\\
&\approx \frac{\sigma_g^2N_p}{I_{tot}^2}\left(1-\frac{N_s}{\sqrt{2\pi}\sigma_g}e^{-\frac{N_s^2}{8\sigma_g^2}}\right)\erf^2\left(\frac{N_s}{2\sqrt{2}\sigma_g}\right)
\label{eq:variance_shot_noise_continuum}
\end{align}
In the limit of very large ROI ($N_s\rightarrow \infty$), Eq.~\ref{eq:variance_shot_noise_continuum} approaches the limit $\sigma_{phot}=\sigma_g/\sqrt{N_p}$, in agreement to the result derived in \cite{winick1986cramer, irwan1999analysis, rousset1999adaptive}. Although Eq.~\ref{eq:variance_shot_noise_continuum} is often a good approximation, it is affected by a typical error of $\approx 10\%$ for small PSF radius of $\sigma_g=0.6$, when the approximation of the discrete sum in Eq.~\ref{eq:variance_shot_noise_sum} by a continuous integral is no longer valid.

\subsection{Variance of the biased estimator}
The contributions of pixel noise and shot noise to the overall centroid error shall be represented by the independent random variables $n_{pix}$ and $n_{phot}$, respectively, and the center position of the incident light relative to the center pixel by the random variable $X_0$ which is uniformly distributed between $[-0.5,0.5]$. Using Eq.~\ref{eq:corrected_estimator} for the relation between centroid $X_c$ and the actual position $X_0$ via function $f_{lu}$, it is convenient to represent the centroid errors in terms of a stochastic equation to find the total CoG variance $\sigma_x^2$:
\begin{align}
X_c&=f_{lu}(X_0)+n_{pix}+n_{phot}\label{eq:stochastic_equation}\\
\sigma_x^2&=\langle (X_c-X_0)^2\rangle=\langle(\delta_{x,tot}^2(X_0)\rangle+\langle n_{pix}^2\rangle+\langle n_{phot}^2\rangle,\label{eq:total_variance}
\end{align}
where the last three terms of Eq.~\ref{eq:total_variance} correspond to the variances of contributions to CoG error from systematic error $\sigma_{sys}^2$, pixel noise $\sigma_{pix}^2$, and photon noise $\sigma_{phot}^2$, which have been specified before in Eqs.~\ref{eq:variance_total_systematic_error}, \ref{eq:variance_pixel_noise}, and \ref{eq:variance_shot_noise_sum}, respectively. Note that for large $\sigma_g$ the total systematic error $\delta_{x,tot}$ of Eq.~\ref{eq:total_systematic} can be linearly approximated by the constant $F_{cut}$ defined in Eq.~\ref{eq:linear_truncation_error} to obtain the variance of the system error as $\sigma_{sys}^2=F_{cut}^2/12$.

\subsection{Variance of unbiased estimators}
\label{sec:unbiased_estimator}
Using the definition of CoG estimator with full systematic error correction of Eq.~\ref{eq:corrected_estimator}, the stochastic Eq.~\ref{eq:stochastic_equation} can be transformed as follows:
\begin{align}
X_{c,ub}&=f^{-1}_{lu}(X_c)=f_{lu}^{-1}(f_{lu}(X_0)+n_{pix}+n_{phot})\nonumber\\
&\approx f_{lu}^{-1}(f_{lu}(X_0))+\frac{\partial}{\partial Z}f_{lu}^{-1}(f_{lu}(X_0))(n_{pix}+n_{phot})\nonumber\\
&=X_0+\frac{1}{\frac{\partial}{\partial X_0}f_{lu}(X_0)}\left(n_{pix}+n_{phot}\right),
\label{eq:approximate_corrected_estimator}
\end{align}
where in the second line a Taylor expansion of the function $f_{lu}(Z)$ around $Z=f_{lu}(X_0)$ up to first order in $n_{pix}+n_{phot}$ was used.
The total variance of the unbiased estimator is then found from Eq.~\ref{eq:approximate_corrected_estimator}, as follows:
\begin{equation}
\sigma_{x,ub}^2=\mathrm{Var}(X_{c,ub}-X_0)=F_{broad} (\sigma_{pix}^2+\sigma_{phot}^2),
\label{eq:variance_unbiased_estimator}
\end{equation}
where the broadening factor $F_{broad}$ is given by:
\begin{equation}
F_{broad}=\int_{-0.5}^{0.5} \left(\frac{\partial}{\partial x_0}f_{lu}(x_0)\right)^{-2} dx_0
\end{equation}
Eq.~\ref{eq:variance_unbiased_estimator} shows that the effect of removing the estimator bias is to effectively increase the variances of pixel and photon noise by a constant factor $F_{broad}$ which can be calculated from the calibration function of Eq.~\ref{eq:sys_err_correction} and shown to be always larger than 1. The analytical expressions for the unbiased estimator variances are one major result of this paper which allows quantifying the effect of bias removal and investigating under which conditions systematic error correction is favorable.\\
In a similar way, the variance of linear error correction can be found. Noting that according to Eq.~\ref{eq:linear_truncation_error} the centroid position $X_c$ in absence of random noise relates to the true center $x_0$ of the intensity distribution through $X_c=x_0(1+F_{cut})$, an improved estimator $X_{c,lin}$ is obtained by dividing the centroid position by the constant factor $1+F_{cut}$.
\begin{align}
X_{c,lin}&=\frac{X_c}{1+F_{cut}}=\frac{(f_{lu}(X_0)+n_{pix}+n_{phot})}{1+F_{cut}}\nonumber\\
\label{eq:approximate_lin_estimator}
\end{align}
The total variance of the estimator with linear bias correction is then be found from Eq.~\ref{eq:approximate_lin_estimator}, \ref{eq:variance_pixel_noise}, and \ref{eq:variance_shot_noise_sum}, as follows:
\begin{align}
\sigma_{x,lin}^2&=\mathrm{Var}(X_{c,lin}-X_0)=\sigma_{res}^2+\frac{(\sigma_{pix}^2+\sigma_{phot}^2)}{(1+F_{cut})^2}\label{eq:variance_lin_estimator}\\
\sigma_{res}^2&=\int_{-0.5}^{0.5} \left(\frac{f_{lu}(x_0)}{1+F_{cut}}-x_0\right)^2 dx_0,\label{eq:residual_systematic_error_variance}
\end{align}
where the residual systematic error variance $\sigma_{res}^2$ was introduced. In the limit of large $\sigma_g$ the linear approximation becomes exact and the residual systematic error vanishes. Noting that $F_{cut}<0$, it is clear that a linear correction of the systematic error effectively increases the variances of pixel and photon noise, as seen before for correction of the total systematic error.

\subsection{Discussing the variances}
\label{sec:discussing_variances}
\begin{figure}
\centerline{\includegraphics[width=0.85\columnwidth]{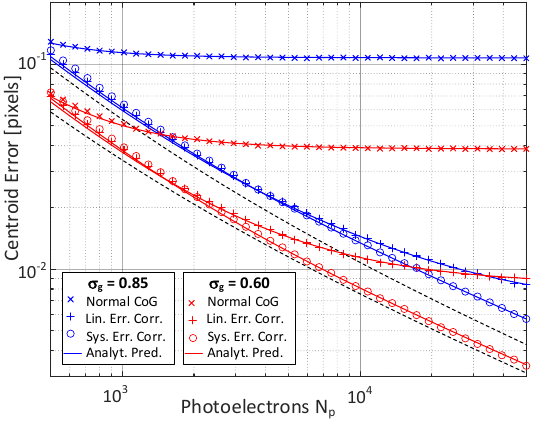}}
\caption  {Centroid error for ROI [3x3] plotted against number of photoelectrons $N_p$ from $500$ to $5\cdot10^4$. Results for PSF radii $\sigma_r=0.85$ and $\sigma_r=0.60$ are colored in blue and red, respectively. Markers designate  MC simulation results for normal CoG errors 'x', linear correction of systematic error  '+', and full correction of systematic error 'o'. Solid colored lines refer to analytical predictions, black dashed lines to the Cramer Rao bounds for $\sigma_g=0.85$ (top) and $\sigma_g=0.60$ (bottom).\label{fig_4}}
\end{figure}

\begin{figure}
\centerline{\includegraphics[width=0.85\columnwidth]{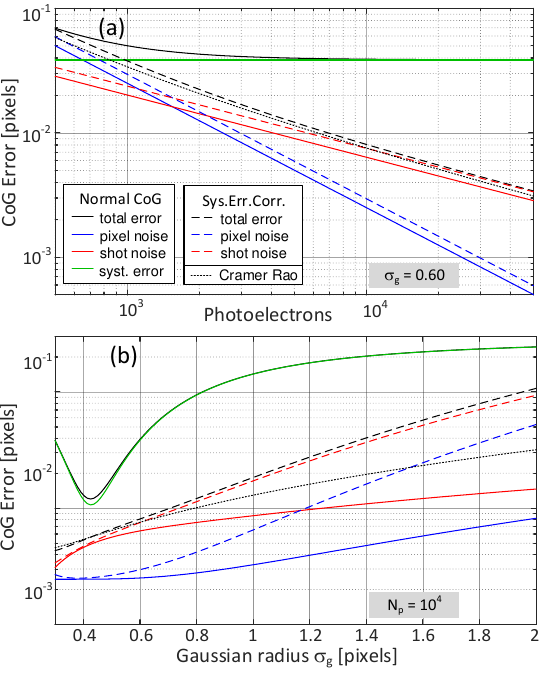}}
\caption  {Contributions of pixel noise (blue), photon noise (red) and systematic error (green) to the total error (black) for normal CoG (solid lines) and CoG with full systematic error correction (dashed lines). Cramer Rao limit is indicated by the black dotted line. Plot of errors against photoelectrons for $\sigma_g=0.6$ (a) and against PSF radius for $N_p=10^4$ (b). ROI size [3x3].\label{fig_5}}
\end{figure}

Fig.~\ref{fig_4} plots the centroid errors against photon number for an ROI of [3x3]. Markers designate results obtained from $2\times 10^4$ MC simulations for normal CoG (marker 'x'), linear systematic error correction (marker '+'), and full systematic error correction (marker 'o'). Data for $\sigma_g=0.85$ and $\sigma_g=0.60$ are colored blue and red, respectively. It is apparent that centroiding errors decrease for both, normal CoG and systematic error correction, when the PSF radius is decreased from $\sigma_g=0.85$ to $\sigma_g=0.60$. In the latter case, the errors of the unbiased estimator (red circles) are very close to the Cramer Rao limit (black dashed line). Solid colored lines refer to analytical error predictions obtained from Eqs.~\ref{eq:total_variance}, \ref{eq:variance_lin_estimator}, and \ref{eq:variance_unbiased_estimator}, respectively. The analytical predictions (colored lines) are found to be in excellent agreement with the MC results (colored markers) for all 3 estimators, which verifies the validity and accuracy of the respective derivations.\\
Fig.~\ref{fig_5} plots the individual contributions of pixel noise (blue), shot noise (red) and systematic error (green) to the total CoG error (black) for the normal CoG estimator $X_c$ of Eq.~\ref{eq:xc_with_noise} (solid lines) and for the unbiased CoG estimator $X_{c,ub}$ with full systematic error correction of Eq.~\ref{eq:approximate_corrected_estimator} (dashed lines). The ROI size is [3x3] and the PSF radius $\sigma_g=0.6$, as was also used for the red curves of Fig.~\ref{fig_4}.  In Fig.~\ref{fig_5}(a), where centroid errors are plotted against the number of photoelectrons $N_p$, the total CoG error (black solid line) is dominated by the systematic error (green solid line) for $N_p>600$ (SNR$>$15), while pixel noise dominates for smaller $N_p$. Shot noise errors become larger than those from pixel noise for $N_p>1500$ (SNR$>$30), defining the asymptote of the total error for the unbiased estimator $X_{ub}$ (black dashed line). While systematic error has been completely removed for $X_{ub}$, the effective contributions of pixel noise (blue dashed line) and shot noise (red dashed line) are increased above their respective values for the biased estimator, as predicted by Eq.~\ref{eq:variance_unbiased_estimator}. For the specific PSF radius used in the plot ($\sigma_g=0.6$), the unbiased estimator performs better than the normal CoG estimator for the plotted range of $N_p$. However, for even smaller numbers of $N_p$, the effect of pixel noise becomes increasingly larger and systematic error correction ineffective, leading to a better performance of the normal CoG estimator. Therefore, for values of the SNR$<$10 no error correction should be performed, while for values of SNR$>$10 the unbiased estimator yields superior performance and ensures that centroiding errors decrease as $\propto 1/SNR$.\\
This is visualized in Fig.~\ref{fig_5}(b) which plots the centroid errors against PSF radius $\sigma_g$ for a large number of photoelectrons $N_p=10^4$ (SNR=95). The unbiased estimator (black dashed line) performs better than the normal CoG estimator for any value of the PSF radius, as the latter is completely dominated by systematic errors (green solid line). Owing to the large number of photoelectrons , the next largest error contribution comes from shot noise (red solid line), which is always larger than the contribution from pixel noise (blue solid line). The effective shot noise (red dashed line) and pixel noise (blue dashed line) of the unbiased estimator are increased with respect to those of the normal CoG estimator by the broadening factor $F_{broad}$ that depends on $\sigma_g$, as predicted by Eq.~\ref{eq:variance_unbiased_estimator}.

\section{Comparison to other centroiding methods}
\label{sec:other_centroiding_methods}
\subsection{Other estimators}
As seen before in Fig.~\ref{fig_3}(a), the normal CoG estimator of Eq.~\ref{eq:xc_with_noise} and the estimators with full and linear systematic error correction of Eqs.~\ref{eq:approximate_corrected_estimator} and Eq.~\ref{eq:approximate_lin_estimator} work best for small ROI of sizes [3x3] and [5x5]. This is because (1) the number of pixels affected by random noise, including dark noise and read-out-noise, is confined to a minimum, and (2) the emergence of systematic errors that increase for smaller ROI sizes is mitigated by correction schemes.\\
Another group of estimators achieves high centroiding accuracy by suppressing noisy pixels with low signal in the centroid computation and ensuring that little bias is introduced through this process. Therefore, these estimators are not required to operate on a small ROI but intrinsically select or weight the pixels most suitable for optimal performance. These estimators include:
\begin{enumerate}
	\item {\bf Thresholded CoG (Thr-CoG)}: Only pixels with a value above a critical threshold $I_{thr}$ are considered in the centroid computation. Typically, a threshold value of $3\sigma_\eta$ above the background level is chosen for best performance \cite{thomas2004optimized, thomas2006comparison, ma2009error}, but it may be favorable to choose a smaller threshold for low light levels.
	\item {\bf Iteratively Weighted CoG (IWCoG)}: The observed pixel intensity distribution is multiplied with a weighting function whose shape is defined by the PSF and whose center position is a variable parameter, before the centroid is computed. The obtained centroid position is then used as center position of the weighting function in the next iteration of the centroid computation. Iterations are made until the change in centroid position between two iterations is below a threshold value \cite{baker2007iteratively, akondi2010improved}. The radius $\sigma_w$ of the weighting function may be used as another variable parameter, but it is typically observed to converge to the value $\sigma_g$ of the PSF after a few iterations \cite{baker2007iteratively}. 
		\item {\bf Iterative Least Squares Fit (ILSQF)}:  A model function is directly fitted to the observed pixel intensity distribution using the iterative Nelder-Mead Simplex Method \cite{lagarias1998convergence}, where the center position and intensity maximum are free fitting parameters. 
		Accuracies were generally better when selecting the pixel-broadened distribution of Eq.~\ref{eq:f_broadened} rather than the intensity profile of Eq.~\ref{eq:Gaussian1d} as model function. Note that the execution time for ILSQF is much longer than that of other algorithms so that it is typically not used in time-critical applications.
		\item {\bf Two-dimensional Gaussian regression (2dGR)}:  This non-iterative method, introduced by Nobach et al.\cite{nobach2005two}, uses a Gaussian filter of kernel [11x11] to filter the image data before performing a Gaussian regression in a small ROI [3x3]. It is much faster than ILSQF.
\end{enumerate}
So far, the centroiding performance has been specified in units of pixel. However, in most applications, the size of the PSF on the detector may be adjusted through the focal length of the imaging system. When increasing the focal length, the PSF radius $\sigma_g$ grows in proportion. The error $\sigma_x$ in its centroid position converts back into an angular error by dividing through the focal length (see Fig.~\ref{fig_1}) which in turn is proportional to $\sigma_g$. Therefore, a fair comparative benchmark for the centroiding performance is given by the normalized centroiding error $\sigma_{n,x}$, defined as the ratio of centroiding error to PSF radius $\sigma_{n,x}=\sigma_x/\sigma_g$, and not by $\sigma_x$ alone. Although this is a key criterion in designing an optimized directional measurement system, it is overlooked in many publications, where comparisons are made only in terms of absolute centroiding error in units of pixels.\\

\subsection{Discussion of simulation results}
\label{sec:other_estimator_simulations}
\begin{figure}
\centerline{\includegraphics[width=0.90\columnwidth]{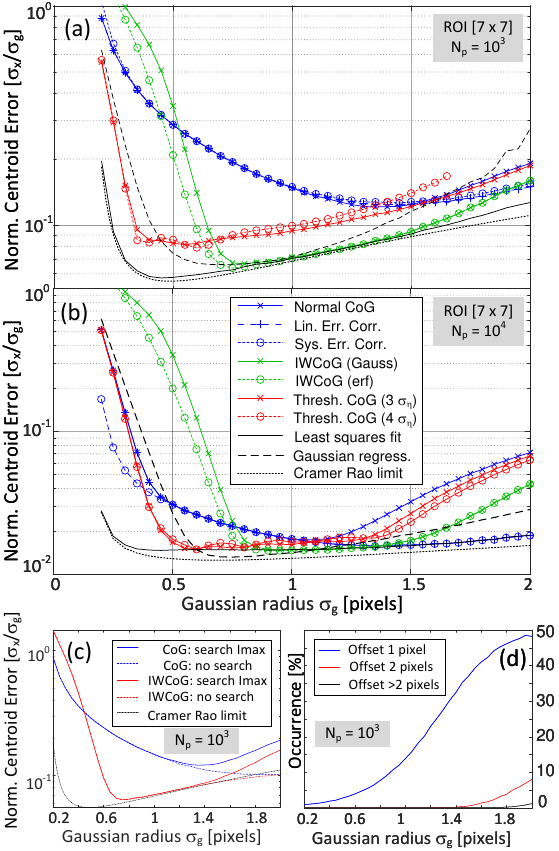}}
\caption  {Normalized centroid error $\sigma_{x,n}$ plotted against PSF radius $\sigma_g$ for $10^3$ (a) and $10^4$ (b) photoelectrons, respectively: Normal CoG (blue x-marks), unbiased CoG (blue plus signs and circles); Thr-CoG with thresholds of $3\sigma_{\eta}$ (red x-marks) and $4\sigma_{\eta}$ (red circles); IWCoG with Gaussian (green x-marks) and pixel-broadened (green circles) weighting function; least-squares fit (black solid line); Gaussian regression (black dashed line); CRLB (black dotted line). (c) $\sigma_{x,n}$ of normal CoG (blue) and IWCoG (red) for simulations with search for intensity maximum (solid lines) and without (dashed lines). $N_p=10^3$. (d) Percentage of ROI offsets, defined as distance of pixel with maximum intensity from pixel where actual center of intensity profile is located: offsets of 1 (blue), 2 (red) and $>2$ (black) pixels. $N_p=10^3$. All plots: ROI [7x7], $N_{MC}=4\times 10^4$. \label{fig_6}}
\end{figure}

In the following, performances of alternative estimators are investigated and compared to those of normal CoG and CoG with systematic error correction.
Figures~\ref{fig_6}(a) and (b) plot the normalized centroiding error $\sigma_{n,x}$ against the PSF radius $\sigma_g$ for an ROI of [7x7] pixels and signal levels of $10^3$ and $10^4$ photoelectrons, respectively. The chosen ROI size [7x7] is sufficiently large for the alternative estimators to perform at their optimum, and increasing the ROI size even more showed no improvement.\\
For small signal, as shown in Fig.~\ref{fig_6}(a), the CoG estimator (blue x-marks) clearly performs worse than the other estimators, and systematic error correction (blue circles) does not improve performance. The thresholded CoG essentially yields the same accuracy for a threshold of $3\sigma_\eta$ (red x-marks) as for $4\sigma_\eta$ (red circles), and a slight oscillatory behavior is observed, as noted in \cite{ares2004influence}. For larger radii, the PSF spreads out more and intensity values at the edges approach the noise floor, resulting in their removal through thresholding. For this reason, the lower threshold performs slightly better at larger PSF radii. The IWCoG (green symbols) performs best of all CoG estimators and approaches the CRLB (black dotted line) for values of $0.75\le\sigma_g\le 1.5$. A slightly better performance is noticeable for the pixel broadened weighting function of Eq.~\ref{eq:intensity_sampling} (green circles) than for the Gaussian weighting function of Eq.~\ref{eq:Gaussian2d} (green x-marks). Gaussian regression (black dashed line) clearly outperforms thresholded CoG and is nearly as good as IWCoG in the range $0.6\le\sigma_g\le 1.0$. The iterative least squares fit (black solid line) is equal or better than other estimators but slightly offset from the CRLB (black dotted line). This may seem surprising, but could be explained by the presence of photon noise whose spatial dependency makes the ILSQF estimator less effective.\\
For large signal $N_p=10^4$, as shown in Fig.~\ref{fig_6}(b), all estimators approach the CRLB, albeit at different values for $\sigma_g$. In this case, CoG with full and linear systematic error correction, indicated by blue circles and plus signs, respectively, both yield significant improvement with respect to the uncorrected CoG estimator (blue x-marks), approaching the CRLB for $\sigma_g\approx 1.3$. The Thr-CoG estimators (red symbols) exhibit best accuracy for small $\sigma_g \approx 0.6$, while the IWCoG estimators (green symbols) require larger PSF radii $\sigma_g\ge 0.8$ for optimal performance. The Gaussian regression (black dashed line) slightly outperforms other estimators in the interval $0.6\le\sigma_g\le 1.0$ because of the application of a Gaussian filter before regression. The centroiding errors of all estimators, as well as the CRLB, rise steeply towards very small PSF radii, where the PSF is greatly undersampled and systematic errors rise abruptly. It is important to note that this transition occurs at much larger PSF radii for the IWCoG than for other estimators. The reason is that for IWCoG centroid computation the pixel intensities are multiplied with an almost identical weighting function, i.e. effectively the squared intensity distribution is used, which reduces $\sigma_g$ by factor $\sqrt{2}$ so that undersampling already occurs for larger PSF radii. This can be remedied, to some extent, by using a broader weighting function, but in general IWCoG should be performed for significantly larger PSF radii than other algorithms.\\
It is important to mention a specific aspect, relating to the difference between target acquisition and tracking. So far, the assumption for the derivation of the analytical expressions of systematic errors in section \ref{sec:unbiased_estimator} has been that the center of the light intensity profile $x_0$ is located within the interval [-0.5,0.5] around the center pixel of the ROI. This is only the case, if the intensity maximum has been correctly identified and the ROI chosen around it accordingly. However, for broad distributions, weak signal, and in the presence of random noise, the pixel of maximum intensity may differ from the one where the distribution center is located, and a ``wrong ROI'' chosen, which increases the centroiding error. The results shown in Figs.~\ref{fig_6}(a) and (b) were obtained after an intensity maximum search was performed for a PSF image located anywhere at random on the detector, as is the case in a realistic ``target acquisition'' scenario. If, however, no such search is performed but $x_0$ is constrained to be within half a pixel distance from the ROI center, the results start differing from those of an acquisition scenario for $\sigma_g \gtrapprox 1.4$ and $N_p=10^3$, as shown by the dashed lines in Fig.~\ref{fig_6}(c) for CoG (blue) and IWCoG (red) estimators. In this case, which could describe a ``target tracking'' scenario with the ROI closely following the target, the centroiding errors for normal CoG surprisingly drop below the CRLB. This is because the CRLB was derived by Winick \cite{winick1986cramer} without assuming any constraints on the location of the spot center on the detector. Fig.~\ref{fig_6}(d) shows that the occurrence of wrongly identified center pixels increases with PSF radius, where offsets of 1 , 2, and more than 2 pixels are represented by blue, red, and black lines, respectively. At the same time that offsets of 2 or more pixels start appearing ($\sigma_g\approx 1.4$), the peak detection threshold, defined in section \ref{sec:noise_sources}, is close to the signal level and centroid errors between the two scenarios diverge from another.\\
Finally, it is interesting to compare the execution times of the various algorithms which are given in Table \ref{table_1} as multiples of the execution time of the normal CoG algorithm. The results were obtained from averaging over 1E4 simulations for an ROI [7 x 7] and for different values of $\sigma_g$ in Matlab. Execution time for pre-processing steps common to all algorithms, i.e. determining the intensity maximum and selecting the ROI around it, is not included.
\begin{table}[t]
\renewcommand{\arraystretch}{1.0}
\caption{Centroiding speed of different algorithms.}
\label{table_1}
\centering
\begin{tabular}{l|l}
\hline
\bfseries Algorithm & rel. exec. time \\
\hline\hline
\bfseries CoG & 1.0\\ 
\bfseries CoG, linear systematic error correction & 1.0 \\
\bfseries CoG, full systematic error correction & 3.4 \\
\bfseries Iteratively weighted CoG & 13.7 \\
\bfseries Thresholded CoG & 2.5 \\
\bfseries 2d Gaussian regression with filter & 16.8 \\
\bfseries Iterative least squares fit (Simplex) & 2595 \\
\end{tabular}
\end{table}
It is apparent that the normal CoG and CoG with linear systematic error correction perform fastest, while CoG with full systematic error correction takes 3.4 times longer as it needs to access a lookup table. Iteratively weighted CoG requires evaluating and applying a weighting function and typically 6 to 8 iterations, making it 13.7 times slower than normal CoG. As mentioned before, Gaussian regression only works well after application of a filter to the image data, which incurs a penalty and makes it 5 times slower than CoG with systematic error correction. An iterative least squares fit using an unconstrained simplex method is most time-consuming and more than two orders of magnitude slower than Gaussian regression. Note that speed benchmarks are dependent on the platform and implementation details so that these figures can only be taken as approximate indicators.\\

\subsection{Optimizing all estimators}
\begin{figure}
\centerline{\includegraphics[width=0.85\columnwidth]{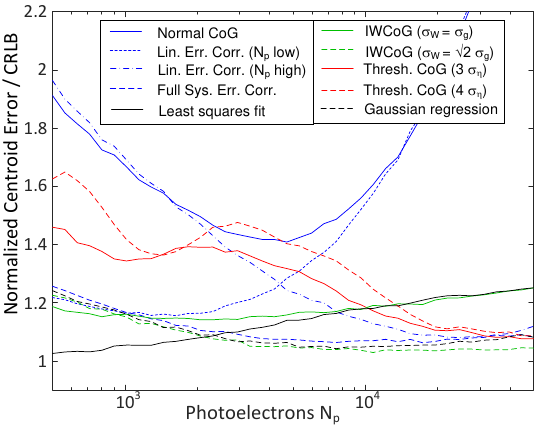}}
\caption  { Normalized centroid error $\sigma_{x,n}$ divided by the Cramer Rao limit for $\sigma_g=0.6$ plotted against $N_p$: Normal CoG (blue solid), CoG with linear systematic error correction for low $N_p$ (blue dotted) and high $N_p$ (blue dash-dotted), CoG with full systematic error correction (blue dashed), thresholded CoG with thresholds $3\sigma_\mu$ (red solid) and $4\sigma_\mu$ (red dashed), IWCoG with $\sigma_w=\sigma_g$ (green solid) and $\sigma_w=\sqrt{2}\sigma_g$ (green dashed), ILSQF (black solid), Gaussian regression (black dashed).\label{fig_7}}
\end{figure}

Table~\ref{table_2} summarizes the results of $N_{MC}=8\times 10^4$ simulations for selected estimators, where two signal regimes of low ($N_p=10^3$) and high ($N_p=10^4$) number of photoelectrons, listed in the upper and lower half of the table respectively, and three different ROI sizes per signal regime were considered. Note that the pixel-broadened weighting function of Eq.~\ref{eq:intensity_sampling} was chosen for the IWCoG , and a threshold of $3\sigma_\mu$ for the Thr-CoG.\\ 
The optimization was performed as follows: For a given signal regime and ROI size, simulated curves such as those plotted in Fig.~\ref{fig_6} were generated for each estimator by incrementing the PSF radius $\sigma_g$ between $0.2$ and $1.5$ in steps of 0.01 pixels. For each estimator, the minimum normalized centroid error $\sigma_{n,x}$ was found, together with the value of $\sigma_g$ where it occurred, and entered into the cells of Table~\ref{table_2} as the first and second number (in brackets), respectively. The numbers in bold indicate the overall minimum of $\sigma_{n,x}$ found among the 3 simulated ROI sizes for a given estimator in the low and the high signal regime.
In the same way, the minimal normalized centroid error and associated PSF radius are determined in the Cramer Rao limit (which does not depend on ROI size) for both signal regimes and listed in the respective header lines for comparison.\\
\begin{table}[t]
\renewcommand{\arraystretch}{1.3}
\caption{Centroiding accuracy for different algorithms.}
\label{table_2}
\centering
\begin{tabular}{l|c|c|c|c}
\hline
\bfseries ROI & \bfseries CoG & \bfseries CoG Corr. & \bfseries IWCoG & \bfseries Thr. CoG \\ 
\bfseries & $\sigma_{n,x}$, $(\sigma_g)$ & $\sigma_{n,x}$, $(\sigma_g)$ & $\sigma_{n,x}$, $(\sigma_g)$ & $\sigma_{n,x}$, $(\sigma_g)$\\
\hline\hline
\multicolumn{5}{c}{$N_p=10^3$, CR limit: $\sigma_{n,x}=0.055$, ($\sigma_g=0.49$)}\\
\hline
\bfseries [3x3]& \textbf{0.074, (0.48)} & \textbf{0.066, (0.60)} & 0.081, (0.69) & \textbf{0.072, (0.53)}\\
\bfseries [5x5]& 0.091, (0.97) & 0.092, (1.01) & 0.064, (0.75) & 0.076, (0.53)\\
\bfseries [7x7]& 0.120, (1.37) & 0.126, (1.37) & \textbf{0.064, (0.78)} & 0.081, (0.51)\\
\hline
\multicolumn{5}{c}{$N_p=10^4$, CR limit: $\sigma_{n,x}=0.013$, ($\sigma_g=0.69$)}\\
\hline
\bfseries [3x3]& 0.028, (0.44) & \textbf{0.013, (0.55)} & 0.049, (0.71) & 0.026, (0.44)\\
\bfseries [5x5]& \textbf{0.015, (0.71)} & 0.014, (0.93) & 0.015, (0.88) & \textbf{0.015, (0.58)}\\
\bfseries [7x7]& 0.017, (1.08) & 0.016, (1.40) & \textbf{0.015, (0.98)} & 0.015, (0.58)\\
\end{tabular}
\end{table}
As expected, the normal CoG estimator works best for small ROIs, with a preference for ROI [3x3] with $\sigma_g=0.48$ for low $N_p$ and ROI [5x5] with $\sigma_g=0.71$ for high $N_p$. In contrast, the CoG estimator with full systematic error correction (second column) performs best for ROI [3x3] in both signal regimes, with similar values of $\sigma_g=0.6$ and $\sigma_g=0.55$, respectively. Surprisingly, it is also the best performing estimator for the high signal regime and the second best for the low signal regime. As already discussed in section \ref{sec:other_estimator_simulations}, the IWCoG estimator requires a large ROI [7x7] and large PSF radii for best performance, with optimal radii of $\sigma_g=0.78$ and $\sigma_g=0.98$ for low and high $N_p$, respectively. Thresholded CoG works best for ROI [3x3] with a small PSF radius of $\sigma_g= 0.53$ for low $N_p$, and ROI [5x5] with $\sigma_g=0.58$ for high $N_p$.\\
It is apparent that the various estimators achieve the lowest values of $\sigma_{x,n}$ for different values of $\sigma_g$ in the two signal regimes, but practical applications typically require good centroiding performance across a range of possible signal levels. Therefore, a compromise must be made and a single value of $\sigma_g$ chosen for which satisfactory performance is obtained in both regimes, and which is then fixed by the appropriate focal length of the imaging system. The optimum configurations in terms of ROI size $N_s$ and PSF radius $\sigma_g$, expressed as tupels $(N_s,\sigma_g)$ for the four estimators of Table ~\ref{table_2}, were selected to be: CoG (5, 0.85), CoG Corr. (3, 0.6), IWCoG (7, 0.85), and Thr. CoG (5, 0.55). 
Additionally, the optimum configurations for four additional estimators, which are not listed in Table \ref{table_2}, were selected to be: iterative least squares estimator (7,0.6), 2d Gaussian regression (3,0.73), linear systematic error correction for low $N_p$ (3,0.6) and for high $N_p$ (5,0.85). \\
The results of simulations across a range of $N_p$ from 500 to $5\times 10^4$ are given in Fig.~\ref{fig_7}, where the normalized centroid errors of the various estimators are divided by the (normalized) CRLB in order to assess by how much any configuration deviates from the theoretical optimum. The CRLB was calculated for the full range of $N_p$ and using $\sigma_g=0.6$.
As expected, the normal CoG (blue solid line) is found to perform worst of all estimators. Linear systematic error correction only works well for either small $N_p$ (dotted blue) or large $N_p$ (dash-dotted blue), surpassing the CoG in the respective regime. Thresholded CoG for threshold of $3\sigma_\mu$ (red solid) is found to be better than for threshold of $4\sigma_\mu$ (red dashed), but significantly worse than IWCoG for low photon number, where its errors are typically 50\% above the CRLB. 
Interestingly, if IWCoG is performed using a radius of the weighting function equal to the PSF radius $\sigma_w=\sigma_g$ (green solid), it performs considerably worse for large $N_p$ than if a broader weighting function of radius $\sigma_w=\sqrt{2}\sigma_g$ (green dashed) is used. 
It is also surprising that the iterative least-squares fit (black solid) is the best estimator only for low $N_p$, but its relative error continuously increases with $N_p$ to over 20\% above the CRLB for $N_p=5\times 10^4$, making it worse than other estimators, in particular considering that the computational effort is much higher. 
Finally, the unbiased CoG with full systematic error correction (blue dashed), the filtered 2d Gaussian regression (black dashed), and the IWCoG with broad weighting function (green dashed) are all found to have the same performance to within a few percent, albeit the CoG with full systematic error correction is the fastest algorithm among the three (see Table \ref{table_1}).

\section{Summary and conclusions}
A comprehensive assessment of systematic errors was made for small ROIs and PSF widths where truncation and sampling errors may become dominant contributors to the overall measurement uncertainty. 
New estimators $X_{c,ub}$ and $X_{c,lin}$ were introduced which perform a correction of the full systematic error and a linear approximation, respectively.
Analytical expressions for the new estimator variances were derived, showing that systematic error can be fully removed but variances from pixel and photon shot noise are slightly increased.
A detailed study of the centroiding performance for small photon number and different ROI sizes indicated that it is favorable to use small ROIs [3x3] and [5x5] for the biased but especially for the unbiased estimators, because small ROIs minimize the accumulated pixel noise.\\ 
The performance of the new estimators with respect to accuracy and computational speed  was compared to the performance of commonly used other estimators, including thresholded CoG, iteratively weighted CoG, least-squares fitting based, and two-dimensional Gaussian regression. Most of these estimators were found to perform more accurately in larger ROIs because of their intrinsic ability to suppress outlying pixels. For all estimators not only the ROI size but in particular also the PSF radius were optimized, revealing that the optimal PSF radius differs between estimators and also changes with photon number.
A direct comparison of all estimators to the Cramer Rao Lower Bound (CRLB), performed for individually optimized configurations, showed that the novel unbiased estimator $X_{c,ub}$ is among the best performing estimators but requires less computational effort than the equally well performing iteratively weighted CoG and 2d Gaussian regression algorithms.

\section*{Acknowledgments}
The author gratefully acknowledges fruitful discussions with T. Lamour, A. Sell, T. Ziegler, and R. Gerndt (Airbus).


\end{document}